\documentclass[epj]{webofc}
\usepackage[varg]{txfonts}   
\usepackage[T1]{fontenc}
\usepackage{mathrsfs}
\usepackage[tbtags]{amsmath}
\usepackage{amssymb}
\usepackage{amsxtra}
\usepackage{amsopn}
\usepackage{latexsym}
\usepackage[mathcal]{eucal}
\usepackage{mathtools}
\wocname{EPJ Web of Conferences}
\woctitle{CONF12}
\newcommand{\BE}{\begin{equation}}
\newcommand{\EE}{\end{equation}}
\newcommand{\BA}{\begin{align}}
\newcommand{\EA}{\end{align}}
\newcommand{\Tr}{\mathrm Tr}
\newcommand{\nn}{\nonumber}

\renewcommand{\Re}{\mathop{\rm Re}}
\renewcommand{\Im}{\mathop{\rm Im}}

\begin{document}
\selectlanguage{english}
\title{The Dark Side of the Propagators: 
exploring their analytic \\properties by a massive expansion}
%
%

\author{Fabio Siringo \inst{1}\fnsep\thanks{\email{fabio.siringo@ct.infn.it}} }

\institute{Dipartimento di Fisica e Astronomia 
dell'Universit\`a di Catania,\\ 
INFN Sezione di Catania,
Via S.Sofia 64, I-95123 Catania, Italy}

\abstract{Analytical functions for the propagators of QCD, including a set of chiral quarks, 
are derived by a one-loop massive expansion in the Landau gauge, and are studied in Minkowski space, 
yielding a direct proof of positivity violation and confinement from first principles. 
Complex conjugated poles are found for the gluon propagator.}
\maketitle
%

\section{introduction}

Most of the non-perturbative approaches to QCD rely on numerical calculations in the
Euclidean space, where a clear picture for the propagators of QCD emerges in Landau gauge
by lattice simulations, by numerical solution of Schwinger-Dyson equations and, more recently,
by unconventional variational methods\cite{sigma,sigma2,gep}. 

However, since physics happens in Minkowski space, many important dynamical information cannot be
extracted by the Euclidean formalism, unless we have an analytic function that can be continued to
the physical space or the whole numerical analysis is carried out in Minkowski space\cite{straussDSE}.
Even the concept of a dynamical mass has no obvious meaning for confined particles
like gluons and quarks. Thus, it is still questioned if the gluon propagator has poles, while
some evidence of positivity violation has only been shown by indirect arguments.

Even if the analytic continuation of a limited set of data points is an ill-defined problem,
a K\"allen-Lehmann spectral function was reconstructed in Ref.\cite{dudal14}
from the lattice data of the gluon propagator, giving some direct evidence for positivity violation and 
the absence of any discrete mass pole on the physical real axis.

Quite recently, an analytical approach has been proposed that is based on a different expansion point for
the exact Lagrangian of pure Yang-Mills theory in the Landau gauge\cite{ptqcd,ptqcd2}. 
The new expansion is around a massive free-particle 
propagator, yielding a {\it massive} loop expansion with massive particles in the internal lines of the
Feynman graphs.  
Moreover, the method does not require too much new effort since most of the required graphs were
evaluated before by other authors\cite{tissier10,tissier11,tissier13,tissier14} in the framework of a 
one-loop phenomenological approach to QCD that is in good agreement with the data of lattice simulations.

From first principles, without adding spurious counterterms or phenomenological parameters, 
at one-loop the expansion provides analytical universal functions for the dressing functions, 
predicting some scaling properties that are satisfied by the data of lattice simulations\cite{scaling}. 
In the Euclidean space and Landau gauge, the massive expansion is in impressive agreement with the lattice data
and the one-loop propagators are analytic functions that can be easily continued and studied in Minkowski space.
Moreover, the massive expansion has been extended to full QCD by the inclusion of a set of
chiral quarks in the Lagrangian\cite{analyt} and 
the dynamical breaking of chiral symmetry is described on the same footing
of gluon mass generation, providing a unified picture from first principles.
Since analytic functions are derived for the one-loop propagators in the Landau gauge, they can be easily
continued to Minkowski space where the spectral functions can be studied in detail.
In this paper, a concise review is given of the main features of the propagators in Minkowski space,
where their {\it dark} side emerges by the optimized massive expansion of Refs.\cite{ptqcd2,analyt}.

\section{The Optimized Massive Expansion}

The full Lagrangian of QCD, including $N_f$ massless chiral quarks, can be written as
\BE
{\cal L}_{QCD}={\cal L}_{YM}+{\cal L}_{fix}+{\cal L}_{FP}+{\cal L}_{q}
\label{LQCD}
\EE
where ${\cal L}_{YM}$ is the Yang-Mills term
\BE
{\cal L}_{YM}=-\frac{1}{2} \Tr\left(  \hat F_{\mu\nu}\hat F^{\mu\nu}\right)
\EE
${\cal L}_{fix}$ is a covariant gauge fixing term, ${\cal L}_{FP}$ is the ghost Lagrangian
arising from the Faddev-Popov determinant and ${\cal L}_{q}$ is the quark Lagrangian
\BE
{\cal L}_{q}=\sum_{i=1}^{N_f}
\bar\Psi_i\left[i
  {\ensuremath{\mathrlap{\>\not{\phantom{a}}}{\>\partial}}}
-g{\ensuremath{\mathrlap{\>\not{\phantom{A}}}A_a}}\hat T_a\right]\Psi_i.
\label{Lq}
\EE

\begin{figure}[b] 
\centering
\includegraphics[width=0.3\textwidth,angle=-90]{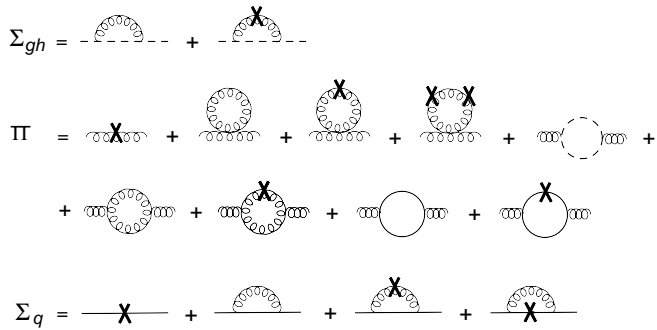}
\caption{Two-point graphs with no more than three vertices and no more than one loop. 
The crosses are the counterterms $\delta \Gamma_g=m^2$, $\delta \Gamma_q=-M$.
In this paper, the quark and ghost self energy and the gluon polarization  
are obtained by the sum of all the graphs in the figure.}
\label{F1}
\end{figure}

The total action is ${\cal S}_{tot}={\cal S}_0+{\cal S}_I$, where the free-particle term 
${\cal S}_0$ is the usual quadratic part that can be written in terms of the standard free-particle propagators
of gluons, quarks and ghosts, namely $\Delta_0$, $S_0$ and ${\cal G}_0$, respectively. 

As shown in Refs.\cite{ptqcd2,analyt} a shift of the pole in the propagators can be introduced by an unconventional
splitting of the total action. We may add and subtract the arbitrary terms $\delta {\cal S}_g$,
$\delta {\cal S}_q$ in the total action
\BE
{\cal S}_0\rightarrow {\cal S}_0+\delta {\cal S}_q+\delta {\cal S}_g,\qquad
{\cal S}_I\rightarrow {\cal S}_I-\delta {\cal S}_q-\delta {\cal S}_g
\label{shift}
\EE
and take
\begin{align}
\delta {\cal S}_g&= \frac{1}{2}\int A_{a\mu}(x)\>\delta_{ab}\> \delta\Gamma_g^{\mu\nu}(x,y)\>
A_{b\nu}(y) {\rm d}^dx{\rm d}^dy \nn\\
\delta {\cal S}_q&=\sum_{i=1}^{N_f}\int
\bar\Psi_i(x)\>\delta\Gamma_q (x,y)\>\Psi_i(y){\rm d}^dx{\rm d}^dy
\label{dS}
\end{align}
where the vertex functions $\delta\Gamma_g$, $\delta\Gamma_q$  are given by
a shift of the inverse propagators
\begin{align}
\delta \Gamma_g^{\mu\nu}(x,y)&=
\left[{\Delta_m^{-1}}^{\mu\nu}(x,y)- {\Delta_0^{-1}}^{\mu\nu}(x,y)\right]\nn\\
\delta \Gamma_q(x,y)&=\left[{S_M^{-1}} (x,y)- {S_0^{-1}} (x,y)\right]
\label{dG}
\end{align}
and ${\Delta_m}^{\mu\nu}$, $S_M$ are massive free-particle propagators 
\begin{align}
{\Delta_m^{-1}}^{\mu\nu} (p)&={\Delta_m}(p)^{-1} t^{\mu\nu}(p)  
+\frac{-p^2}{\xi}\ell^{\mu\nu}(p)\nn\\
{\Delta_m}(p)^{-1}&=-p^2+m^2,\qquad S_M (p)^{-1}=
{\ensuremath{\mathrlap{\>\not{\phantom{p}}}{~p}}}
-M.
\label{Deltam}
\end{align}
Here $t^{\mu\nu}$, $l^{\mu\nu}$ are Lorentz projectors and the  masses $m$ and $M$ are totally arbitrary. 
Since $\delta {\cal S}_q$ and $\delta {\cal S}_g$ 
are added and subtracted again, the total action cannot depend on the masses, but any expansion in powers 
of the new shifted interaction ${\cal S}_I \to {\cal S}_I-\delta {\cal S}_q-\delta {\cal S}_g$ is going 
to depend on them at any finite order because of the truncation. 
Thus, while we are not changing the content of the  theory, the emerging perturbative approximation
is going to depend on the masses and can be optimized by a choice of $m$ and $M$ that minimizes the
effects of higher orders, yielding a variational tool disguised to look like a perturbative method\cite{ptqcd2,analyt}.
The idea is not new and goes back to the works on the Gaussian effective
potential\cite{stevenson,su2LR,HT,stancu2,stancu,superc1,superc2,AF,varlight,bubble}
where an unknown mass parameter was inserted in the zeroth order propagator and subtracted
from the interaction, yielding a pure variational approximation
with the mass that acts as a variational parameter.

The shifts $\delta {\cal S}_q$, $\delta {\cal S}_g$ have two effects on the resulting perturbative expansion: 
the free-particle propagators are replaced by massive propagators and  new two-point vertices 
are added to the interaction, arising from the counterterms that read 
\BE
\delta \Gamma_g ^{\mu\nu}(p)= m^2 t^{\mu \nu} (p), \qquad \delta \Gamma_q (p)= -M.
\EE

The Landau gauge is the optimal choice for the massive expansion\cite{ptqcd2} and from now on we will 
take the limit $\xi\to 0$. In Eq.(\ref{Deltam}) the gluon propagator becomes transverse 
and we can simplify the notation and drop the projectors $t^{\mu\nu}$ everywhere whenever each term 
is transverse. Moreover we drop all color indices in the diagonal matrices.

We can use the standard formalism of Feynman graphs with massive zeroth
order propagators $\Delta_m$, $S_M$ and the counterterms $\delta \Gamma_g=m^2$, $\delta \Gamma_q=-M$
that must be added to the standard vertices of QCD. 

Assuming that the effective coupling never reaches values that are too large\cite{ptqcd2}, 
we may neglect higher loops and take a double expansion in powers of the total 
interaction and in powers of the coupling, retaining graphs with $n$ vertices
at most and no more than $\ell$ loops. 

The graphs contributing to the quark and ghost self-energy and to the gluon polarization are shown
in Fig.~\ref{F1} up to the third order and one-loop. Their calculation is straightforward and explicit
analytical expressions are reported in Refs.\cite{ptqcd2,analyt}.

\section{Analytic continuation: pure Yang-Mills theory}

The dressed propagators of pure SU(N) Yang-Mills theory
can be written as
\BE
\Delta(p)^{-1}=-p^2+\frac{5}{8}\alpha m^2-\left[ \Pi(p)-\Pi(0)\right],\qquad
{\cal G} (p)^{-1}=p^2-\Sigma_{gh} (p)
\label{dressed}
\EE
where the ghost self-energy $\Sigma_{gh}$ and the gluon polarization $\Pi$
were evaluated in Ref.\cite{ptqcd2} as a sum of the graphs in Fig.~\ref{F1} 
(omitting quark loops) and $\alpha$ is an effective coupling.
The one-loop gluon and ghost propagators are made finite by
standard wave function renormalization and explicit analytic expressions
were derived by dimensional regularization in Ref.\cite{ptqcd2}.

\begin{figure}[b] 
\centering
\includegraphics[width=0.45\textwidth,angle=-90]{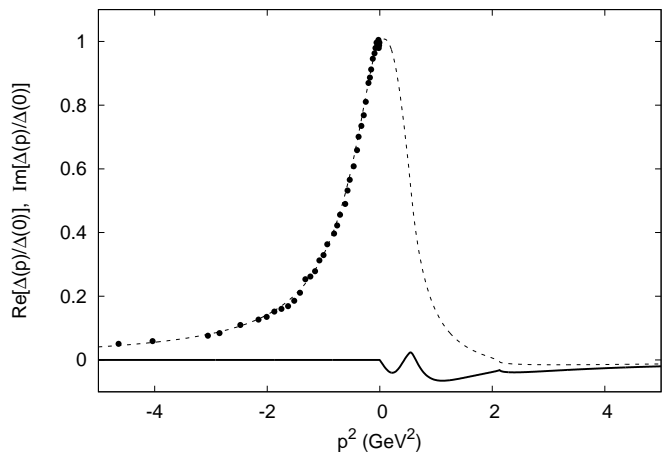}
\caption{The real and the imaginary part  of the gluon propagator
are displayed together with the lattice data of Ref.\cite{bogolubsky} ($N=3$, $\beta=5.7$, $L=96$). 
The propagator is normalized by its finite value at $p^2=0$ and is evaluated by Eq.(\ref{dress3})
with the optimal choice $F_0=-1.05$ and $m=0.73$ GeV.}
\label{F2}
\end{figure}

\begin{figure}[t] 
\centering
\includegraphics[width=0.45\textwidth,angle=-90]{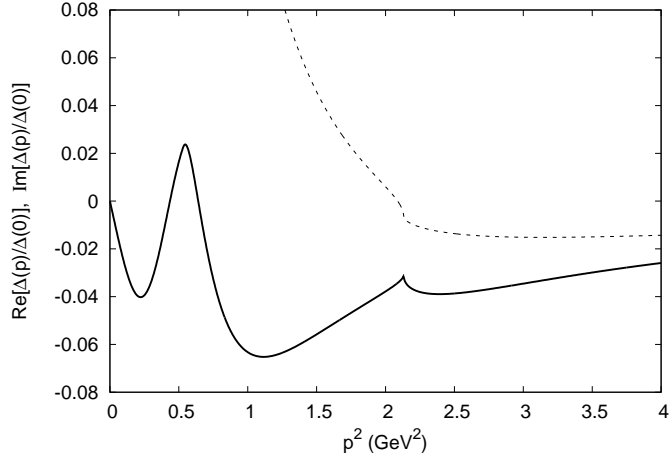}
\caption{The real and the imaginary part of the gluon propagator
(enlargement of Fig.~\ref{F2}).}
\label{F4}
\end{figure}

\begin{figure}[t] 
\centering
\includegraphics[width=0.45\textwidth,angle=-90]{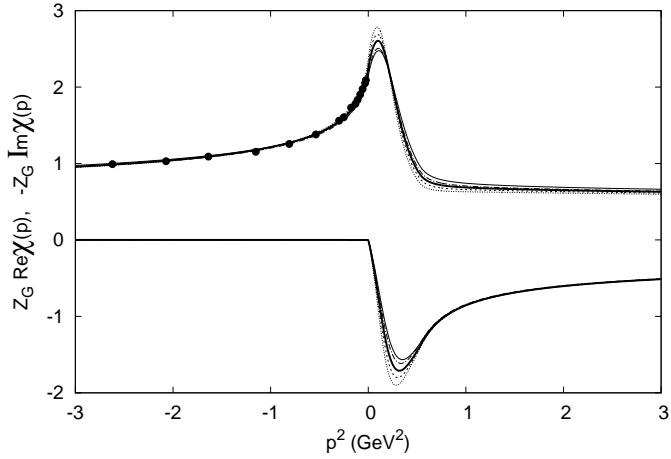}
\caption{Real part $\Re\chi$ and imaginary part $-\Im\chi=\pi p^2\rho$ of the ghost dressing function
according to Eq.(\ref{dress3}) for $m=0.73$ GeV and
several values of $G_0$ in the range $0.2<G_0<0.3$. The points are the lattice data 
of Ref.\cite{bogolubsky} ($N=3$, $\beta=5.7$, $L=80$). The best agreement with the data points is 
obtained for $G_0=0.24$ (solid line). The dressing function is scaled by a finite
renormalization constant $Z_G$.}
\label{F5}
\end{figure}

\begin{figure}[t] 
\centering
\includegraphics[width=0.55\textwidth,angle=-90]{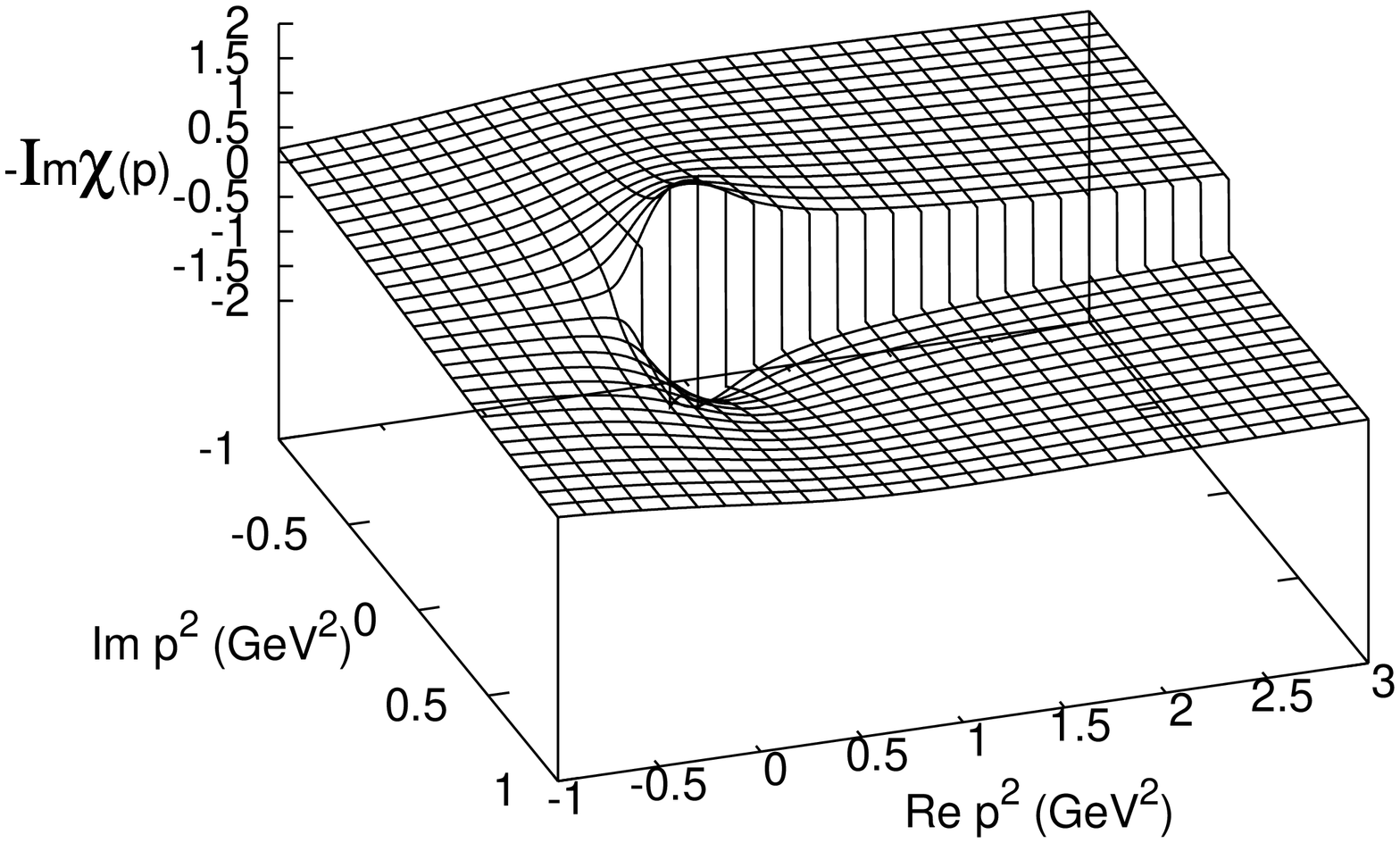}
\caption{Imaginary part $-\Im\chi=\pi p^2\rho$ of the ghost dressing function
in the complex plane for $m=0.73$ GeV and $G_0=0.24$}
\label{GH3D}
\end{figure}

It is useful to introduce the adimensional ghost and gluon dressing functions 
\BE
\chi (p)=p^2{\cal G}(p),\quad J(p)=-p^2 \Delta (p).
\label{dress1}
\EE

They can be written as
\BE
\left[\alpha\> \chi (s)\right]^{-1}=G(s)+G_0\quad
\left[\alpha\> J (s)\right]^{-1}=F(s)+F_0
\label{dress3}
\EE
where $s=-p^2/m^2$ is the Euclidean momentum and the
two adimensional functions $F(s)$, $G(s)$ are given by the polarization and self energy graphs
in Fig.~\ref{F1}, while all the constants are grouped together in the finite one-loop
renormalization constants $F_0$ and $G_0$ that are the only free parameters to be optimized.
Being equivalent to a variation of the subtraction point, any change of the additive constant can
be seen as a variation of the renormalization scheme yielding a special case of 
{\it optimized perturbation theory} that has been proven to be very effective for the convergence 
of the expansion\cite{stevensonRS16}.

A very important consequence of Eq.(\ref{dress3}) is that, up to an arbitrary
{\it multiplicative} renormalization constant, the inverse dressing functions are given by
the universal functions $F(s)$ and $G(s)$ up to an {\it additive} renormalization constant.
Such scaling property is satisfied quite
well by the lattice data for SU(2) and SU(3) that collapse on the same universal curves $F(s)$, $G(s)$ in the
infrared\cite{ptqcd,ptqcd2,scaling,analyt}, thus confirming that higher order terms can
be made negligible by an optimized choice of the constants $F_0$, $G_0$. 

For $SU(3)$ and $-p^2<4$ GeV$\>^2$ the lattice data of Ref.\cite{bogolubsky} are very well reproduced 
by setting $F_0=-1.05$ and $m=0.73$ GeV in Eq.(\ref{dress3}). 
Some deviation occurs for $-p^2> 4$ GeV$\>^2$ because of the large logs that
require a resummation by RG equations in the UV.

The gluon propagator can be continued to Minkowski space by setting
$s=-p^2/m^2-i\varepsilon$ and the resulting complex function is shown in Fig.~\ref{F2}.
The imaginary part  has a cut for $p^2>0$ where it defines a spectral function.
The lack of any sharp peak or pole on the real axis and the violation of positivity can
be regarded as a direct proof of confinement.

Out of the real axis, in the complex plane, the propagator has two conjugated poles at 
$p^2\approx (0.16\pm 0.60i)$ GeV$^2$, close to the imaginary axis, as predicted
by the {\it i-particle} scenario\cite{iparticle} emerging from 
the refined version\cite{dudal08,dudal08b,dudal11}
of the Gribov-Zwanziger model\cite{GZ}.

The one-loop ghost propagator, by Eq.(\ref{dress3}) mantains a pole at $p^2=0$. The analytic
continuation $s=-p^2/m^2-i\varepsilon$  yields
\begin{align}
\Re {\cal G}(p^2+i\varepsilon)&=\frac{\Re \chi (p^2)}{p^2}\nn\\
\Im {\cal G}(p^2+i\varepsilon)&=\frac{\Im \chi (p^2)}{p^2}-\pi\chi(0)\>\delta(p^2)
\label{RIG}
\end{align}
and we can define a spectral function on the cut
\BE
\rho(p^2)=-\frac{1}{\pi}\Im {\cal G}(p^2+i\varepsilon)=
\chi(0)\>\delta(p^2)-\frac{1}{\pi}\frac{\Im \chi(p^2)}{p^2}
\label{rho2}
\EE
which has a continuous term given by the imaginary part of the dressing function divided
by $-p^2$.  The details of the continuous term of the spectral function are shown in Fig.~\ref{F5}
by the direct plot of $-\Im \chi$ , together with the real part $\Re \chi$ and the lattice
data of Ref.\cite{bogolubsky} ($N=3$, $\beta=5.7$, $L=80$). We observe that the discrete and the continuous
terms have opposite sign in Eq.(\ref{rho2}), violating the positivity condition. In the Euclidean range
$p^2<0$, the ghost dressing function is not too much sensitive to a change of the additive constant $G_0$. 
In Fig.~\ref{F5}, a change of $G_0$ in the range $0.2<G_0<0.3$ is compensated by a change of the finite 
renormalization constant $Z_G$, so that $Z_G\chi(p^2)$ stays on the lattice data points. 
The best agreement is found for $G_0=0.24$ and is shown as a solid line in Fig.~\ref{F5}. 
The imaginary part has a wide peak at $p^2\approx (0.56)^2$ GeV$\>^2$ and never changes sign.
As shown in Fig.~\ref{GH3D}, it is finite in the whole complex plane, 
with a cut on the real axis where the spectral function is defined by Eq.~(\ref{rho2}).

\section{Analytic continuation: chiral QCD}

The inclusion of a set of chiral quarks requires the calculation of the quark loops contributing to
the gluon polarization and the quark self-energy $\Sigma_q$ as shown in Fig.~\ref{F1}.
Since there are no one-loop graphs with quark lines that contribute to the ghost self-energy $\Sigma_{gh}$,
the one-loop ghost dressing function of QCD is the same of pure Yang-Mills theory.

The gluon polarization of the full theory is obtained from the result for pure Yang-Mills
theory by just adding the quark loops of Fig.~\ref{F1}. Explicit analytical expressions are
reported in Ref.\cite{analyt}.

\begin{figure}[t] 
\centering
\includegraphics[width=0.45\textwidth,angle=-90]{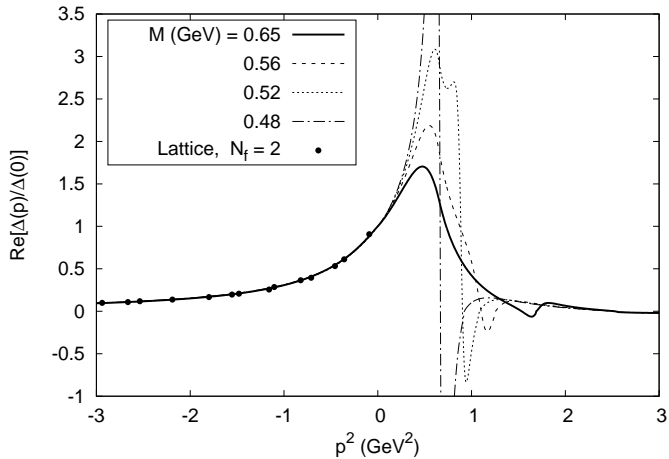}
\caption{The real part of the gluon propagator is evaluated by setting $s=-p^2/m^2-i\varepsilon$,
for $m=0.80$ GeV and several values of $M=0.48,0.52,0.56,0.65$ GeV. 
The constant $F_0$ varies in
the range $-0.65<F_0<-0.6$ in order to keep all curves on the lattice data in the Euclidean
space, for $p^2<0$. The data points are extracted from Fig.~1 of Ref.\cite{binosi12} for $N_f=2$.
The propagator is normalized by its finite value at $p^2=0$.}
\label{B2}
\end{figure}

\begin{figure}[t] 
\centering
\includegraphics[width=0.45\textwidth,angle=-90]{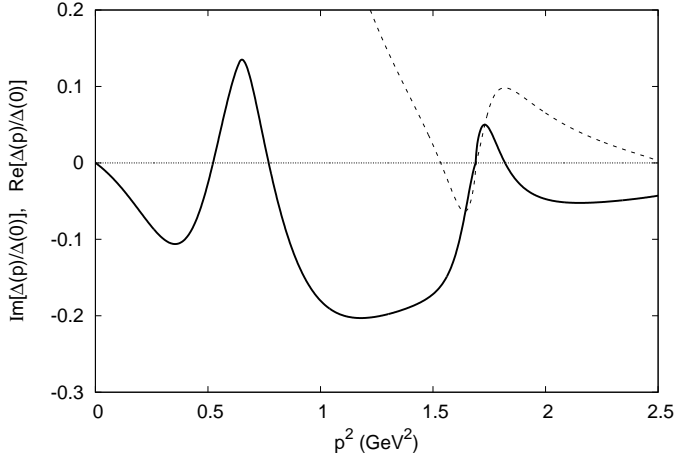}
\caption{The imaginary part of the propagator is evaluated by setting $s=-p^2/m^2-i\varepsilon$
for the optimal set $m=0.80$ GeV, $M=0.65$ GeV,  $F_0=-0.65$ (solid line).
The dashed line is a detail of the real part. The propagator is normalized by its finite value at $p^2=0$.}
\label{B4}
\end{figure}

The real part of the gluon propagator is shown in Fig.~\ref{B2} for $s=-p^2/m^2-i\varepsilon$. While rather
insensitive to the choice of $M$ in the Euclidean space, the shape of the propagator depends 
on $M$ when plotted as a function of the time-like momentum $p^2>0$. 
The data points in the figure are the lattice data of Ref.\cite{binosi12} for two light
quarks, having no lattice data for the gluon propagator in the chiral limit.
We observe the presence of a 
positive peak at $p^2\approx m^2$ and a negative peak just before the two-particle threshold $p^2\approx (2M)^2$
where the real part of the propagator changes sign and becomes positive. 
As shown in Fig.~\ref{B4}, where the
imaginary part of the propagator is displayed, the spectral density becomes negative
and its positivity violation is a direct proof of confinement.
At variance with pure
Yang-Mills theory, there is a two-particle threshold at $p^2\approx (2M)^2$ where
the spectral function turns positive for a while. 
Besides being more rich on the real positive axis $p^2>0$, for $N_f=2$ the unquenched gluon propagator has
more poles in the complex plane. For the optimal set $m=0.8$ GeV, $M=0.65$ GeV We find two pairs of 
conjugated poles at $p^2\approx (1.69,\pm 0.1)$ GeV$^2$ and $p^2\approx (0.54,\pm 0.52)$ GeV$^2$.

The quark self energy $\Sigma_q$ is evaluated by the tree term 
$\delta \Gamma_q=-M$ and 
the three one-loop graphs in Fig.~\ref{F1}. 
In the dressed quark propagator $S(p)$
the mass $M$ is canceled by the tree term $\delta \Gamma_q=-M$.
However, even in the chiral limit,
a mass function is generated for the quarks by the interaction
terms\cite{analyt}.

The dressed quark propagator can be written as
\BE
S(p)= S_p (p^2)
{\ensuremath{\mathrlap{\>\not{\phantom{p}}}{~p}}}
+S_M (p^2)
\label{SPM}
\EE
where the scalar functions $S_p$, $S_M$ follow from the one-loop self energy, yielding
explicit analytical expressions that can be easily continued to Minkowski space by setting
$s=-p^2/m^2-i\varepsilon$.
The imaginary parts have a cut on the real positive axis $p^2>0$ where we can define two spectral densities
\begin{align}
\rho_M(p^2)&=-\frac{1}{\pi}\Im S_M(p^2)\nn\\
\rho_p(p^2)&=-\frac{1}{\pi}\Im S_p(p^2)
\label{rhoq}
\end{align}
so that the propagator reads
\BE
S(p)=\int_0^\infty {\rm d} q^2\frac{\rho_p(q^2) 
{\ensuremath{\mathrlap{\>\not{\phantom{p}}}{~p}}}
+ \rho_M (q^2)}{p^2-q^2+i\varepsilon}.
\label{spectral}
\EE
Any observable fermion must satisfy the positivity conditions
\BE
\rho_p(p^2)\geq 0
\label{cond1}
\EE
\BE
p\> \rho_p(p^2)-\rho_M(p^2)\geq 0
\label{cond2}
\EE
that are strongly violated by the quark propagator, yielding a direct proof of confinement. 

\begin{figure}[!ht] 
\centering
\includegraphics[width=0.45\textwidth,angle=-90]{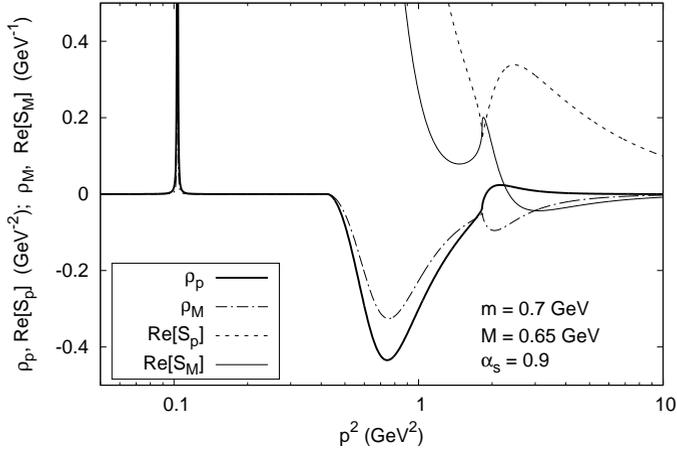}
\caption{Details of the quark spectral functions for $\alpha_s=0.9$, $M=0.65$ GeV, $m=0.7$ GeV.}
\label{C10}
\end{figure}

\begin{figure}[t] 
\centering
\includegraphics[width=0.45\textwidth,angle=-90]{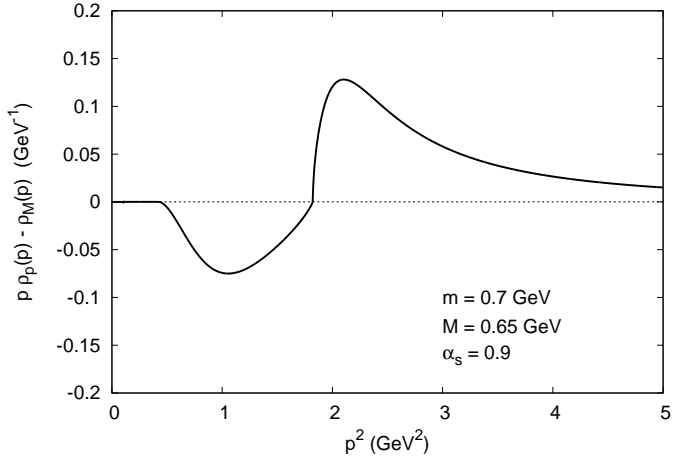}
\caption{The quark spectral function $[p\>\rho_p (p^2)-\rho_M(p^2)]$ is shown as a function of
the physical momentum $p^2=-p_E^2$ for $m=0.7$ GeV, $\alpha_s=0.9$ and $M=0.65$ GeV. 
The positivity condition of Eq.(\ref{cond2}) is violated for $p^2>q_1^2\approx M^2$ below the two-particle
threshold $q^2_2\approx (m+M)^2$.}
\label{C11}
\end{figure}

The spectral functions are shown in Figs.~\ref{C10}, \ref{C11} 
for $\alpha_s=0.9$, $M=0.65$ GeV. We recognize
a discrete term at $p\approx 0.32$ GeV, that arises from the pole of the
propagator.  We can identify two different thresholds.
A first threshold $q_1^2\approx M^2$ at the onset of a {\it negative} continuum spectral density 
($q_1^2\approx (0.65)^2\approx 0.42$ GeV$^2$ in Figs.~\ref{C10}, \ref{C11}).
A second threshold $q_2^2\approx (M+m)^2$ where the spectral density turns positive
($q_2^2\approx (1.35)^2\approx 1.82$ GeV$^2$ in Figs.~\ref{C10}, \ref{C11}). 
While this second threshold can be identified
with the usual two-particle threshold and the high-energy states have a
positive spectral density above $q_2\approx (m+M)$, the negative spectral density above $q_1\approx M$ has 
no obvious physical meaning. It violates the positivity condition (\ref{cond1}) and 
cannot be related to any kind of free-particle
behavior. Thus the quark propagator can only describe confined particles.
No complex poles are observed for the quark propagator.

\section{Concluding remarks}

The massive expansion that was developed for pure Yang-Mills
theory in Refs.\cite{ptqcd,ptqcd2} and extended to full QCD in Ref.\cite{analyt} has been reviewed 
and used as a tool for exploring the dark side of the propagators in Minkowski space.

By a direct comparison with the lattice data,
the expansion is optimized in the Euclidean space yielding accurate analytic propagators that can
be easily continued to Minkowski space. Thus the method provides a powerful tool for the study of 
dynamical properties and spectral functions that can  be hardly extracted from any numerical
data set. From this point of view, the massive expansion is very predictive and gives
a direct proof of positivity violation and confinement for all the particles involved. 

While no direct dynamical content can be given to the gluon mass and to the mass parameters, the
discrete one-particle term in the quark spectral functions can be identified as the (confined) 
physical mass of the constituent quarks.
On the other hand, the mass parameters $m$, $M$ are strongly related to the thresholds of the
spectral functions and determine their rich behavior that is observed in Minkowski space.


\begin{thebibliography} {99}
\bibitem{sigma} F. Siringo and L. Marotta, Eur. Phys. J. C {\bf  44}, 293 (2005).
\bibitem{sigma2} F. Siringo, Mod. Phys. Lett. A {\bf 29}, 1450026 (2014), [arXiv:1308.4037]. 
\bibitem{gep}  F. Siringo, Phys. Rev. D {\bf 88}, 056020 (2013); Phys. Rev. D {\bf 89},  025005 (2014); 
Phys. Rev. D {\bf  90}, 094021 (2014); Phys. Rev. D {\bf 92}, 074034 (2015); arXiv:1507.05543.
\bibitem{straussDSE} S. Strauss, C. S. Fischer, C. Kellermann, Phys. Rev. Lett. {\bf 109}, 252001 (2012).
\bibitem{dudal14} D. Dudal, O. Oliveira, P. J. Silva, Phys. Rev. D {\bf 89}, 014010 (2014).
\bibitem{ptqcd}  F. Siringo, arXiv:1509.05891.
\bibitem{ptqcd2} F. Siringo, Nucl. Phys. B {\bf 907}, 572 (2016), [arXiv:1511.01015].
\bibitem{tissier10} M. Tissier, N. Wschebor, Phys. Rev. D {\bf 82}, 101701(R) (2010).
\bibitem{tissier11} M. Tissier, N. Wschebor, Phys. Rev. D {\bf 84}, 045018 (2011).
\bibitem{tissier13} U. Reinosa, J. Serreau, M. Tissier, N. Wschebor, Phys. Rev. D {\bf 89}, 105016 (2014).
\bibitem{tissier14} M. Pelaez, M. Tissier, N. Wschebor, Phys. Rev. D {\bf 90}, 065031 (2014).
\bibitem{scaling} F.Siringo, EPJ Web of Conferences 137, 13016 (2017), [arXiv:1607.02040].
\bibitem{analyt} F. Siringo, Phys. Rev. D {\bf 94}, 114036 (2016), [arXiv:1605.07357].

\bibitem{stevenson} P.M. Stevenson, Phys. Rev. D {\bf 32}, 1389 (1985).
\bibitem{su2LR} F. Siringo, L. Marotta, Phys. Rev. D {\bf 78}, 016003 (2008); Phys. Rev. D {\bf 74}, 115001 (2006).
\bibitem{HT} F. Siringo, Phys. Rev. D {\bf 86}, 076016 (2012), [arXiv: 1208.3592v2].
\bibitem{stancu2} I. Stancu and P. M. Stevenson, Phys. Rev. D {\bf 42}, 2710 (1990).
\bibitem{stancu} I. Stancu, Phys. Rev. D {\bf 43}, 1283 (1991).
\bibitem{superc1} M. Camarda, G.G.N. Angilella, R. Pucci, F. Siringo, 
Eur. Phys. J. B {\bf 33}, 273 (2003).
\bibitem{superc2} L. Marotta, M. Camarda, G.G.N. Angilella and F. Siringo, 
Phys. Rev. B {\bf 73}, 104517 (2006).             
\bibitem{AF} L. Marotta and F. Siringo, Mod. Phys. Lett. B, {\bf 26}, 1250130 (2012).
\bibitem{varlight} F. Siringo,  Phys. Rev. D {\bf 62}, 116009 (2000); Europhys. Lett. {\bf 59}, 820 (2002).
\bibitem{bubble} F. Siringo and L. Marotta, Int. J. Mod. Phys. A {\bf 25}, 5865 (2010).
\bibitem{stevensonRS16} P.M. Stevenson, Nucl. Phys. B {\bf 868}, 38 (2013); Nucl. Phys. B {\bf 910}, 469 (2016).
\bibitem{bogolubsky} I.L. Bogolubsky, E.M. Ilgenfritz, M. Muller-Preussker, A. Sternbeckc, 
Phys. Lett. B {\bf 676}, 69 (2009).
\bibitem{iparticle} L. Baulieu, D. Dudal, M. S. Guimaraes, M. Q. Huber,
S. P. Sorella, N. Vandersickel, D. Zwanziger, Phys.Rev.D {\bf 82}, 025021 (2010).
\bibitem{dudal08} D. Dudal, J. A. Gracey, S. P. Sorella, N. Vandersickel, H. Verschelde, Phys. Rev. D {\bf 78},
065047 (2008).
\bibitem{dudal08b} D.Dudal, S.P.Sorella, N.Vandersickel, H.Verschelde, Phys. Rev. D {\bf 77}, 071501 (2008).
\bibitem{dudal11} D. Dudal, S. P. Sorella, N. Vandersickel, Phys. Rev. D {\bf 84}, 065039 (2011).
\bibitem{GZ} D. Zwanziger, Nucl. Phys. B {\bf 323}, 513 (1989).
\bibitem{binosi12} A. Ayala, A. Bashir, D. Binosi, M. Cristoforetti and J. Rodriguez-Quintero, 
Phys. Rev. D {\bf 86}, 074512 (2012).

\end{thebibliography}
\end{document}